\documentclass[useAMS,usenatbib,usegraphicx]{mn2e}
\usepackage{pdflscape}
\usepackage{graphicx}
\usepackage{longtable}
\usepackage{lscape}
\usepackage{amssymb}
\usepackage{endnotes} 
\usepackage{footnote}
\usepackage{subfigure}
\usepackage{txfonts}
\usepackage{natbib}
\usepackage{url}
\usepackage{helvet}
\usepackage{setspace}

\bibpunct{(}{)}{;}{a}{,}{,}

\def\apjl{{ApJ}}%
\def\apjs{{ApJS}}%
\def\aap{{A\&A}}%
\title[SN~2012aw: Polarimetry]{Broad band polarimetric follow-up of Type IIP SN 2012aw}
\author[Brajesh Kumar et al.]{Brajesh Kumar$^{1,2}$ \thanks{E-mail:
brajesh@aries.res.in, brajesharies@gmail.com}, 
{S. B. Pandey}$^{1}$,
{C. Eswaraiah}$^{1,3}$,
{J. Gorosabel}$^{4,5,6}$ \\
$^{1}$Aryabhatta Research Institute of Observational Sciences, Manora Peak,
Nainital 263 002, India \\
$^{2}$ Institut d'Astrophysique et de G\'{e}ophysique, Universit\'{e} de
Li\`{e}ge, All\'{e}e du 6 Ao\^{u}t 17, B\^{a}t B5C, 4000 Li\`{e}ge, Belgium \\
$^{3}$ Institute of Astronomy, National Central University, 300 Jhongda Rd, Jhongli, 
Taoyuan Country 32054, Taiwan \\
$^{4}$ Instituto de Astrof\'{\i}sica de Andaluc\'{\i}a (CSIC),  Glorieta de la 
Astronom\'{\i}a s/n, 18008 Granada, Spain \\
$^{5}$ Unidad Asociada Grupo Ciencia Planetarias UPV/EHU-IAA/CSIC, Departamento de F\'{\i}sica Aplicada I, \\ 
E.T.S. Ingenier\'{\i}a, Universidad del Pa\'{\i}s Vasco UPV/EHU, Alameda de Urquijo s/n, E-48013 Bilbao, Spain \\
$^{6}$ Ikerbasque, Basque Foundation for Science, Alameda de Urquijo 36-5, E-48008 
Bilbao, Spain.
}
\begin{document}
\date{Accepted ------------, Received ------------; in original form ------------}
\pagerange{\pageref{firstpage}--\pageref{lastpage}} \pubyear{}
\maketitle
\label{firstpage}
\begin{abstract}

We present the results based on $R$-band polarimetric follow-up observations of the nearby
($\sim$10 Mpc) Type II-plateau SN~2012aw. Starting from $\sim$10 days after the SN 
explosion, these polarimetric observations cover $\sim$90 days (during the plateau phase) 
and are distributed over 9 epochs. To characterize the Milky Way interstellar polarization 
(ISP$_{\rm MW}$), we have observed 14 field stars lying in a radius of 10$^\circ$ around 
the SN. We have also tried to subtract the host galaxy dust polarization component assuming 
that the dust properties in the host galaxy are similar to that observed for Galactic dust 
and the general magnetic field follow the large scale structure of the spiral arms of a galaxy.
After correcting the ISP$_{\rm MW}$, our analysis infer that SN~2012aw has maximum 
polarization of 0.85\% $\pm$ 0.08\% but polarization angle does not show much variation 
with a weighted mean value of $\sim$138$\degr$. However, if both ISP$_{\rm MW}$ and host 
galaxy polarization (ISP$_{\rm HG}$) components are subtracted from the observed 
polarization values of SN, maximum polarization of the SN becomes 0.68\% $\pm$ 0.08\%. 
The distribution of $Q$ and $U$ parameters appears to follow a loop like structure.
The evolution of polarimetric light curve (PLC) properties of this event is also compared 
with other well studied core-collapse supernovae of similar type. 

\end{abstract}


\section{Introduction}\label{sec:introduction}

Core-collapse supernovae (CCSNe) exhibit significant level of polarization during various phases 
of their evolution at optical/infra-red wavelengths. In general, the degree of polarization of 
different types of SNe seems to increase with decreasing mass of the stellar envelope at the time 
of explosion \citep[see][]{2000AIPC..522..445W, 2001ApJ...553..861L, 2001ApJ...550.1030W, 
2005ASPC..342..330L}. Type II SNe are polarized at level of $\sim$1\% $-$ 1.5\%. However, Type Ib/c 
SNe (also known as stripped-envelope SNe as the outer envelopes of hydrogen and/or helium of their 
progenitors are partially or completely removed before the explosion) demonstrate significantly 
higher polarization in comparison to Type II SNe \citep[for more details, see][and references therein]
{2001PASP..113..920L, 2002ApJ...580L..39K, 2003ApJ...593L..19K, 
2003ApJ...591.1110W, 2006A&A...459L..33G, 2007ApJ...671.1944M, 2012A&A...545A...7P, 2012ApJ...754...63T, 
2013MNRAS.433L..20M}. The higher polarization values observed in case of Type Ib/c SNe most probably 
arise due to extreme departure from the spherical symmetry \citep{1992SvAL...18..168C, 2001AIPC..586..459H, 
2001AIPC..556..301K}. 

Theoretical modelling predicts that in general CCSNe show the degree of asymmetry of the order of 
10\% $-$ 30\% if modelled in terms of oblate/prolate spheroids \citep[e.g.][]{1991A&A...246..481H}. 
Numerical simulations \citep[see][]{2006ApJ...651..366K, 2011MNRAS.410.1739D} indicate that in case of 
Type II SNe, the level of polarization is also influenced by SN structure (e.g., density and ionization), 
apart from their initial mass and rotation. The possible progenitors of Type IIP SNe are low-mass red/blue 
super-giants and their polarization studies are extremely useful to understand the SN structure in detail.
In spite of being the most common subtypes among the known CCSNe, polarization studies of Type IIP SNe 
have been done only in a handful of cases \citep[e.g.][]{1988MNRAS.234..937B, 2001ApJ...553..861L, 
2006Natur.440..505L, 2001PASP..113..920L, 2006AstL...32..739C, 2010ApJ...713.1363C, 2012AIPC.1429..204L}.
In general, intrinsic polarization in these SNe are observed below 1\% but few exceptions exist in 
literature (for example \citet{2010ApJ...713.1363C} reported $\sim$1.5\% for SN~2006ov).

Systematic polarimetric studies have been started, only after the observations of Type IIP 
SN~1987A \citep[see][]{1988MNRAS.231..695C, 1988ApJ...334..295M, 1991ApJS...77..405J}.
\citet{1982ApJ...263..902S} first pointed out that polarimetry provides direct powerful probe 
to understand the SN geometry \citep[see also][]{1984MNRAS.210..829M, 1991A&A...246..481H}. 
Polarization is believed to be produced due to electron scattering within the SN ejecta. 
When light passes through the expanding ejecta of CCSNe, it retains information about the 
orientation of layers. In spherically symmetric scenario, the equally present directional 
components of the electric vectors will be canceled out to produce zero net polarization. 
If the source is aspherical, incomplete cancellation occurs which finally imprint a net 
polarization (see Fig. 1 of \citealt{2004cetd.conf...30F} and \citealt{2005ASPC..342..330L}). 
In addition to asphericity of the electron scattering atmosphere, there are several other 
processes which can produce polarization in CCSNe such as scattering by dust 
\citep[e.g.][]{1996ApJ...462L..27W}; clumpy ejecta or asymmetrically distributed radioactive 
material within the SN envelope \citep[e.g.][]{1995ApJ...440..821H, 2006AstL...32..739C}, 
and aspherical ionization produced by hard X$-$rays from the interaction between the SN shock 
front and a non-spherical progenitor wind \citep{1996ssr..conf..241W}.

To diagnose the underlying polarization in SNe explosions, two basic techniques i.e. 
broad-band polarimetry and spectropolarimetry have been used. Both of these techniques 
have advantages and disadvantages relative to each other. One of the main advantages 
of spectropolarimetry of SNe with respect to broad-band polarimetry is its ability to infer 
geometric and dynamical information for the different chemical constituents of the explosion. 
Broad-band polarimetric observations construct a rather rough picture of the stellar death 
but require lesser number of total photons than spectropolarimetry. Hence broad-band 
polarimetric observations can be extended to objects at higher red-shifts or/and they 
allow to enhance the polarimetric coverage and sampling of the light curve (LC), especially 
at epochs far from the maximum when the SN is dimmer.

The scope of this paper uses imaging polarimetric observations in $R$-band using 
a meter class telescope when the SN~2012aw was bright enough ($R$ $<$ 13.20 magnitude).

\subsection{SN~2012aw}

SN~2012aw was discovered in a face-on ($i$ $\sim$54.6$^\circ$, from 
HyperLEDA\footnote{http://leda.univ-lyon1.fr - \citet{2003A&A...412...45P}}), 
barred and ringed spiral galaxy M95 (NGC 3351) by P. Fagotti on CCD images taken on 2012 
March 16.85 UT with a 0.5-m reflector \citep[cf. CBET 3054,][]{2012CBET.3054....1F}. The SN was 
located 60\arcsec\, west and 115\arcsec\, north of the center of the host galaxy with coordinates 
$\alpha = 10^{\rm h} 43^{\rm m} 53\fs73$, $\delta =+11\degr 40\arcmin 17\farcs9$ (equinox 2000.0). 
This SN discovery was also confirmed independently by A. Dimai on 2012 March 16.84 UT, and J. 
Skvarc on March 17.90 UT \citetext{more information available in \citealt{2012CBET.3054....1F}, 
CBET 3054; see also special notice no. 269 available at 
AAVSO\footnote{http://www.aavso.org/aavso-special-notice-269}}. 
The spectra obtained on March 17.77 UT by \citet*{2012CBET.3054....3M} with the Asiago Observatory 
1.22-m reflector showed a very blue continuum, essentially featureless, with no absorption bands 
and no detectable emission lines. In subsequent spectra taken on March 19.85 UT 
\citep*{2012CBET.3054....2I} and 19.92 UT \citep{2012CBET.3054....4S}, the line characteristics 
finally led to classify it as a young Type II-P supernova. The explosion date of this event is 
precisely determined by \citet{2012ApJ...759L..13F} and \citet{2013MNRAS.433.1871B}. We adopt 
2012 March 16.1 $\pm$ 0.8 day (JD 2456002.6 $\pm$ 0.8, taken from the later study) as time of 
explosion throughout this article. At a distance of about 10 Mpc 
\citep[cf.][]{2001ApJ...553...47F, 2002ApJ...565..681R, 2013MNRAS.433.1871B}, 
this event provided us a good opportunity to study its detail polarimetric properties. 

\begin{figure}
\centering
\includegraphics[scale = 0.11]{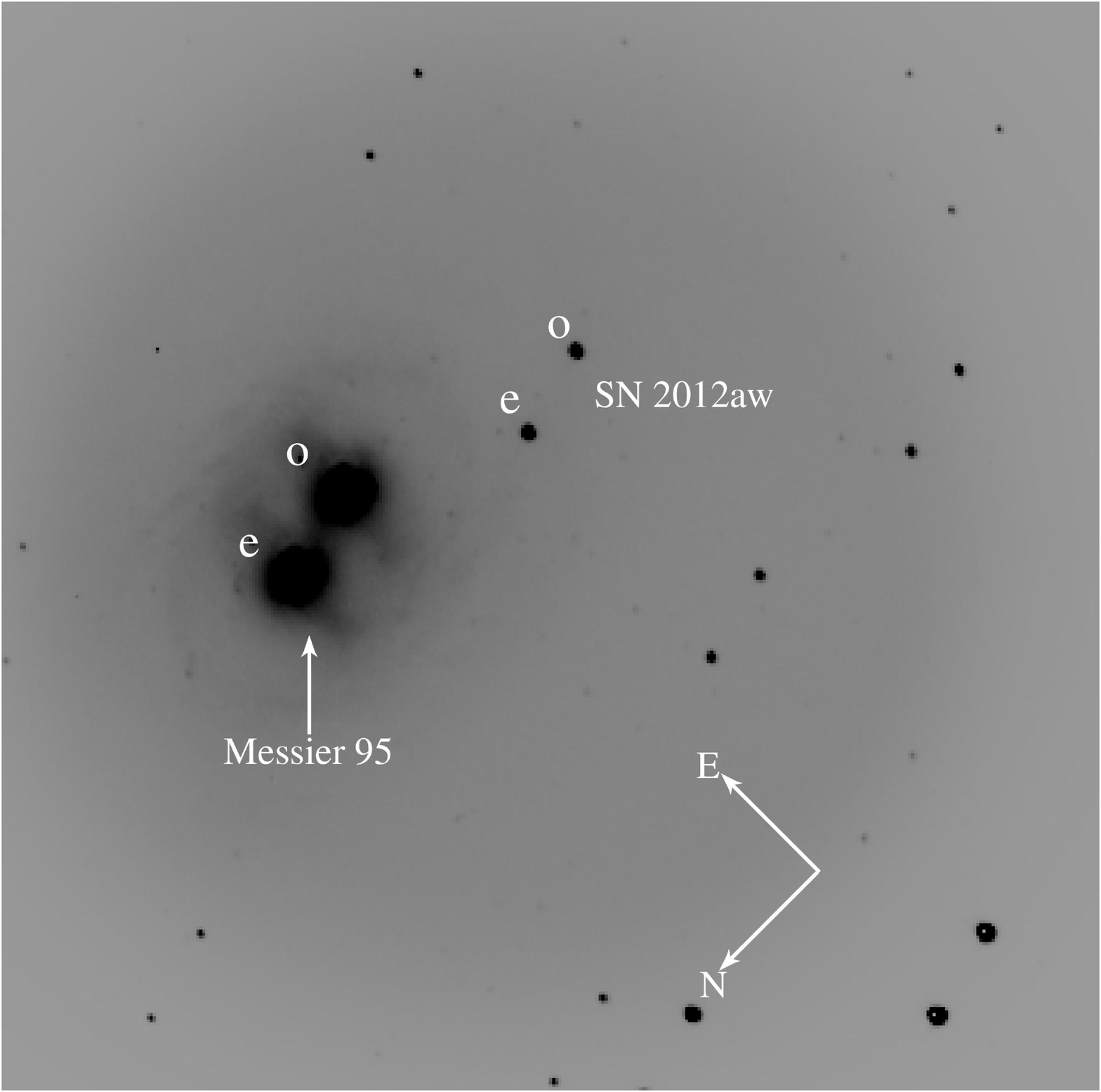}
\caption{
The $R$-band image of the SN~2012aw field around the host galaxy M95,
observed on 17 April 2012 using AIMPOL with the 1-m ST, India. Each object has
two images. The ordinary and extra-ordinary images of SN~2012aw and host galaxy
are labeled as o and e, respectively. The galaxy is marked with a white arrow and
the SN is located 60\arcsec\ west, 115\arcsec\ south of the center of M95 galaxy.
North and East directions are also indicated.}
\label{sn12aw_field}
\end{figure}

\begin{table*}
\centering
\caption{Polarimetric observation log and estimated polarimetric parameters of SN~2012aw.
\label{sn2012aw_log}}
\begin{tabular}{ccc|cc|cc|cc}
\hline \hline
UT Date &JD      & Phase$^{a}$& \multicolumn{2}{c}{Observed}  & \multicolumn{2}{c}{Intrinsic (ISP$_{\rm MW}$} subtracted)  & \multicolumn{2}{c}{Intrinsic (ISP$_{\rm MW}$ + ISP$_{\rm HG}$} subtracted) \\
(2012)  &2450000 &(Days)      & $P_{R} \pm \sigma_{P_{R}}$    & $ \theta{_R} \pm \sigma_{\theta{_R}}$ & $P_{R} \pm \sigma_{P_{R}}$ & $\theta{_R} \pm \sigma_{\theta{_R}}$ & $P_{R} \pm \sigma_{P_{R}}$ & $\theta{_R} \pm \sigma_{\theta{_R}}$ \\
        &        &            & ($\%$)            &($^\circ)$ &   ($\%$) &  ($^\circ$)  &   ($\%$) &  ($^\circ$)                 \\
\hline
Mar 26 &6013.35& 10.75 &0.58 $\pm$ 0.46& 131.4 $\pm$ 22.9& 0.61 $\pm$ 0.46& 138.9 $\pm$ 21.6& 0.39 $\pm$ 0.46& 134.2 $\pm$ 33.4\\
Mar 28 &6015.23& 12.63 &0.56 $\pm$ 0.03& 132.0 $\pm$  1.5& 0.60 $\pm$ 0.03& 139.6 $\pm$  1.4& 0.38 $\pm$ 0.03& 135.1 $\pm$  2.2\\
Mar 29 &6016.28& 13.68 &0.49 $\pm$ 0.08& 132.2 $\pm$  4.6& 0.53 $\pm$ 0.08& 140.8 $\pm$  4.3& 0.31 $\pm$ 0.08& 136.1 $\pm$  7.3\\
Apr 16 &6034.18& 31.58 &0.24 $\pm$ 0.17& 132.0 $\pm$ 21.0& 0.30 $\pm$ 0.17& 147.8 $\pm$ 16.7& 0.07 $\pm$ 0.17& 150.5 $\pm$ 72.8\\
Apr 17 &6035.25& 32.65 &0.26 $\pm$ 0.01& 142.6 $\pm$  1.0& 0.36 $\pm$ 0.01& 154.0 $\pm$  0.8& 0.15 $\pm$ 0.01& 164.8 $\pm$  1.8\\
May 15 &6063.05& 60.45 &0.87 $\pm$ 0.08& 123.8 $\pm$  2.6& 0.85 $\pm$ 0.08& 129.0 $\pm$  2.6& 0.68 $\pm$ 0.08& 123.3 $\pm$  3.3\\
May 19 &6067.04& 64.44 &0.54 $\pm$ 0.01& 124.3 $\pm$  0.5& 0.54 $\pm$ 0.01& 132.7 $\pm$  0.5& 0.35 $\pm$ 0.01& 123.6 $\pm$  0.8\\
May 21 &6069.08& 66.48 &0.43 $\pm$ 0.06& 112.3 $\pm$  4.0& 0.37 $\pm$ 0.06& 122.7 $\pm$  4.6& 0.28 $\pm$ 0.06& 103.4 $\pm$  6.2\\
Jun 14 &6093.23& 90.63 &0.47 $\pm$ 0.14& 128.2 $\pm$  8.5& 0.49 $\pm$ 0.14& 137.5 $\pm$  8.2& 0.29 $\pm$ 0.14& 129.9 $\pm$ 14.1\\
\hline
\end{tabular}  \\
$^{a}$ with reference to the explosion epoch JD 2456002.6\\
\end{table*}

The progenitor of this SN has been detected both in ground and space based pre-explosion 
images and its distinct characteristics are analyzed. In pre-SN explosion images obtained with 
{\it HST}\footnote{Hubble Space Telescope} + WFPC2\footnote{Wide-Field and Planetary Camera 2}, 
VLT\footnote{Very Large Telescope} + ISAAC\footnote{Infrared Spectrometer And Array Camera} and 
NTT\footnote{New Technology Telescope}+SOFI\footnote{Infrared spectrograph and imaging camera},
\citet{2012ApJ...759L..13F} found that the progenitor is a red super-giant 
(mass 14$-$26 M$_{\sun}$). An independent study by \citet{2012ApJ...756..131V} confirmed 
these findings (mass 15$-$20 M$_{\sun}$). However, \citet*{2012ApJ...759...20K} have a 
different view and have concluded that progenitor mass in earlier studies are significantly 
overestimated and that the progenitor's mass is $\textless$ 15 M$_{\sun}$.    
Immediately after the discovery, several groups have started the follow-up observations of this event 
in different wavelengths \citep[see, e.g.][]{2012ATel.3995....1I, 2012ATel.4012....1S, 
2013ApJ...764L..13B, 2013NewA...20...30M, 2013arXiv1311.3568Y}.
Early epoch (4 to 270 days) low-resolution optical spectroscopic and dense photometric 
follow-up (in $UBVRI$/$griz$ bands) observations of SN~2012aw has been analyzed by 
\citet{2013MNRAS.433.1871B}.
In a recent study, \citet{2013arXiv1311.2031J}, have presented nebular phase (between 250 $-$ 451 days) 
optical and near-infrared spectra of this event and have analyzed it with spectral model calculations.
Furthermore, the preliminary analysis of optical spectropolarimetric data of SN~2012aw, revealed that 
outer ejecta is substantially asymmetric \citep{2012ATel.4033....1L}.

In this paper, we present Cousins $R$-band polarimetric follow-up observations of SN~2012aw.
The observations and data reduction procedures are presented in Section~\ref{sec:obs_analy}. 
Estimation of intrinsic polarization is described in Section~\ref{sec:results_discussion}. 
Finally, results and conclusions are presented in Sections \ref{diss} and \ref{sec:conclusions},
respectively. 


\begin{table*}
\centering
\caption{Observational detail of 14 isolated field stars selected to subtract the interstellar polarization.
Observations of all field stars were performed on 20 January 2013 in $R$ band with the 1.04 m ST. All these
stars were selected with known distances and within 10$\degr$ radius around SN~2012aw. The distance
mentioned in the last column has been taken from \citet{2007A&A...474..653V} catalogue.
\label{tab:field_stars}}
\begin{tabular}{lcccll}
\hline \hline
Star    &  RA (J2000) & Dec (J2000)  & $P_{R} \pm \sigma_{P_{R}}$ & $ \theta{_R} \pm \sigma_{\theta{_R}}$ &  Distance  \\
id      & ($^\circ$)  & ($^\circ$)   & $\%$                       & ($^\circ$)                            &  (in pc )  \\
\hline
HD 99028$^{\dagger}$ &    170.98071&     +10.52960 &  0.08 $\pm$ 0.00 & 167.9 $\pm$  1.7 &  23.7  $\pm$   0.5  \\
HD 88830$^{\dagger}$ &    153.73935&     +09.21180 &  0.10 $\pm$ 0.01 & 116.8 $\pm$  1.8 &  36.3  $\pm$   3.8  \\
HD 87739$^{\dagger}$ &    151.78235&     +08.76970 &  0.05 $\pm$ 0.01 &  99.9 $\pm$  6.6 &  85.0  $\pm$   8.3  \\
HD 97907$^{\dagger}$ &    168.96624&     +13.30750 &  0.17 $\pm$ 0.05 &  59.6 $\pm$  9.5 &  99.6  $\pm$  12.1  \\
HD 88282$^{\dagger}$ &    152.72730&     +07.69460 &  0.12 $\pm$ 0.01 &  79.1 $\pm$  1.8 & 118.5  $\pm$  10.0  \\
HD 87635$^{\dagger}$ &    151.57707&     +07.94470 &  0.17 $\pm$ 0.00 &  89.0 $\pm$  0.5 & 135.7  $\pm$  19.9  \\
HD 87915$^{\dagger}$ &    152.08824&     +07.57300 &  0.11 $\pm$ 0.01 &  86.4 $\pm$  1.6 & 193.1  $\pm$  34.7  \\
HD 87996$^{\dagger}$ &    152.20123&     +06.71740 &  0.20 $\pm$ 0.04 &  62.5 $\pm$  5.6 & 243.3  $\pm$  91.2  \\
HD 88514$^{\dagger}$ &    153.15102&     +07.67730 &  0.18 $\pm$ 0.03 &  90.5 $\pm$  4.5 & 254.5  $\pm$  82.9  \\
G 452                &    160.45186&     +12.10886 &  0.10 $\pm$ 0.01 &  22.6 $\pm$  2.4 & 261.1  $\pm$  70.9  \\
BD+12 2250           &    161.08996&     +11.33560 &  0.12 $\pm$ 0.08 & 100.1 $\pm$ 18.0 & 286.5  $\pm$  91.1  \\
BD+13 2299           &    161.41026&     +12.46724 &  0.20 $\pm$ 0.00 &  72.4 $\pm$  0.8 & 314.5  $\pm$  87.0  \\
HD 93329             &    161.65268&     +11.18412 &  0.12 $\pm$ 0.03 & 144.8 $\pm$  5.8 & 358.4  $\pm$ 118.2  \\
HD 92457             &    160.15550&     +12.07868 &  0.05 $\pm$ 0.07 &  27.8 $\pm$ 41.3 & 460.8  $\pm$ 191.1  \\
\hline
\end{tabular}  \\
$^{\dagger}$ Stars with available $V$-band polarimetry from \citet{2000AJ....119..923H} catalogue. \\
BD+12 2250, BD+13 229, G 452, HD 93329 and HD 92457 are the stars within 2$\degr$ radius field 
around the SN. \\
\end{table*}

\section{Observations and data reduction}\label{sec:obs_analy}

Polarimetric observations of SN~2012aw field were carried out during nine nights, i.e., 
26, 28, 29 March; 16, 17 April; 15, 19, 21 May and 12 June 2012 using the ARIES Imaging 
Polarimeter \citep[AIMPOL,][]{2004BASI...32..159R} mounted at the Cassegrain focus of 
the 104-cm Sampurnanand telescope (ST) at Manora Peak, Nainital. This telescope is operated 
by the Aryabhatta Research Institute of Observational sciences (ARIES), India. Complete log 
of these observations is presented in Table \ref{sn2012aw_log}. The position of SN, which 
is fairly isolated from the host galaxy and lies on a smooth and faint galaxy background is 
shown in Fig.~\ref{sn12aw_field}. The observations were carried out in $R$ ($\lambda_{R_{eff}}$ 
= 0.67$\mu$m) photometric band using liquid nitrogen cooled Tektronix 1024 $\times$ 1024 
pixel$^2$ CCD camera. Each pixel of the CCD corresponds to 1.73 arcsec and the 
field-of-view (FOV) is $\sim$8 arcmin in diameter on the sky. The full width at half-maximum 
of the stellar images vary from 2 to 3 pixel. The readout noise and the gain of the CCD are 
7.0 $e^{-}$ and 11.98 $e^{-}$/ADU respectively.

The AIMPOL consists of a half-wave plate modulator and a Wollaston prism beam-splitter.
In order to obtain the measurements with good signal-to-noise ratio, images that 
were acquired at each position of half-wave plate were combined.
Since AIMPOL is not equipped with a narrow-window mask, care was taken to 
exclude the stars that have contamination from the overlap of ordinary and extraordinary
images of one star on the same of another star in the FOV. 

Fluxes of ordinary ($I_{o}$) and extra-ordinary ($I_{e}$) beams of the SN and field stars 
with good signal-to-noise ratio were extracted by standard aperture photometry after preprocessing 
using the {\small IRAF}\footnote{{\small IRAF} is the Image Reduction and Analysis 
Facility distributed by the National Optical Astronomy Observatories, which are operated by the 
Association of Universities for Research in Astronomy, Inc., under cooperative agreement with 
the National Science Foundation.} package. The ratio $R(\alpha)$ is given by: 
\begin{equation}\label{R_alpha}
R(\alpha) = \frac{\frac{{I_{e}}(\alpha)}{{I_{o}}(\alpha)}-1} {\frac{I_{e}(\alpha)} {I_{o}(\alpha)}+1} 
= P \cos(2\theta - 4\alpha),
\end{equation}
Where, $P$ is the fraction of the total linearly polarized light and, $\theta$ is the polarization 
angle of the plane of polarization. Here $\alpha$ is the position of the fast axis of the half-wave 
plate at 0$\degr$, 22.5$\degr$, 45$\degr$ and 67.5$\degr$ corresponding to the four normalized Stokes 
parameters respectively, $q$ [R(0$\degr$)], $u$ [R(22.5$\degr$)], $q_{1}$ [R(45$\degr$)] and 
$u_{1}$ [R(67.5$\degr$)]. The detailed procedures used to estimate the polarization and 
polarization angles for the programme stars are described by \citet{1998A&AS..128..369R, 
2004BASI...32..159R} and \citet{2010MNRAS.403.1577M}. 
Since polarization accuracy is, in principle, limited by photon statistics,
we estimated the errors in normalized Stokes parameters
$\sigma_{R(\alpha)}$ ($\sigma_{q}$, $\sigma_{u}$, $\sigma_{q_{1}}$ and
$\sigma_{u_{1}}$ in $\%$) using the expression \citep{1998A&AS..128..369R}:
\begin{equation}\label{stok_err}
\sigma_{R(\alpha)}=\sqrt{(N_{e}+N_{o}+2N_{b})}/(N_{e}+N_{o}) \\
\end{equation} 
Where, $N_{e}$ and $N_{o}$ are the counts in extra-ordinary and ordinary rays 
respectively, and $N_{b}[=\frac{N_{be}+N_{bo}}{2}]$ is the average background counts 
around the extra-ordinary and ordinary rays of a source. 
The individual errors associated with the four values of $R(\alpha)$, estimated using equation 
(\ref{stok_err}), are used as weights in the calculation of $P$ and $\theta$ for the programme stars.


To correct the measurements for the instrumental polarization and the zero-point polarization 
angle, we observed a number of unpolarized and polarized standards, respectively, taken from 
\cite{1992AJ....104.1563S}. Measurements for the standard stars are compared with those 
taken from the \cite{1992AJ....104.1563S}. The observed values of degree of polarization ($P (\%) $) and 
position angle ($\theta (^\circ)$) were in good agreement (within the 
observational errors) with those published in \cite{1992AJ....104.1563S}. The instrumental polarization 
of AIMPOL on the 1.04-m ST has been characterized and monitored since 2004 for different projects 
and found to be $\sim$0.1\% in different bands \citep[e.g.,][and references therein]{2004BASI...32..159R, 
2009MNRAS.396.1004P, 2011MNRAS.411.1418E, 2012MNRAS.419.2587E, 2013A&A...556A..65E}.

\section{Estimation of intrinsic polarization}\label{sec:results_discussion}

The observed polarization measurements of a distant SN could be composed of various 
components such as interstellar polarization due to Milky Way dust (ISP$_{\rm MW}$),
interstellar polarization due to host galactic dust (ISP$_{\rm HG}$) and due to 
instrumental polarization. As described 
in the previous section, we have already subtracted the instrumental polarization. Therefore, now 
it is essential to estimate the contributions due to ISP$_{\rm MW}$ and ISP$_{\rm HG}$, 
and to remove them from the total observed polarization measurements of SN. However, 
there is no totally reliable method to derive the ISP$_{\rm MW}$/ISP$_{\rm HG}$ of SN 
polarimetry observationally and utmost careful analysis is required to avoid any possible 
fictitious result. In the following sections, we discuss in detail about the 
ISP$_{\rm MW}$ and ISP$_{\rm HG}$ estimation in the present set of observations.


\begin{figure*}
\begin{centering}
\includegraphics[scale = 0.57]{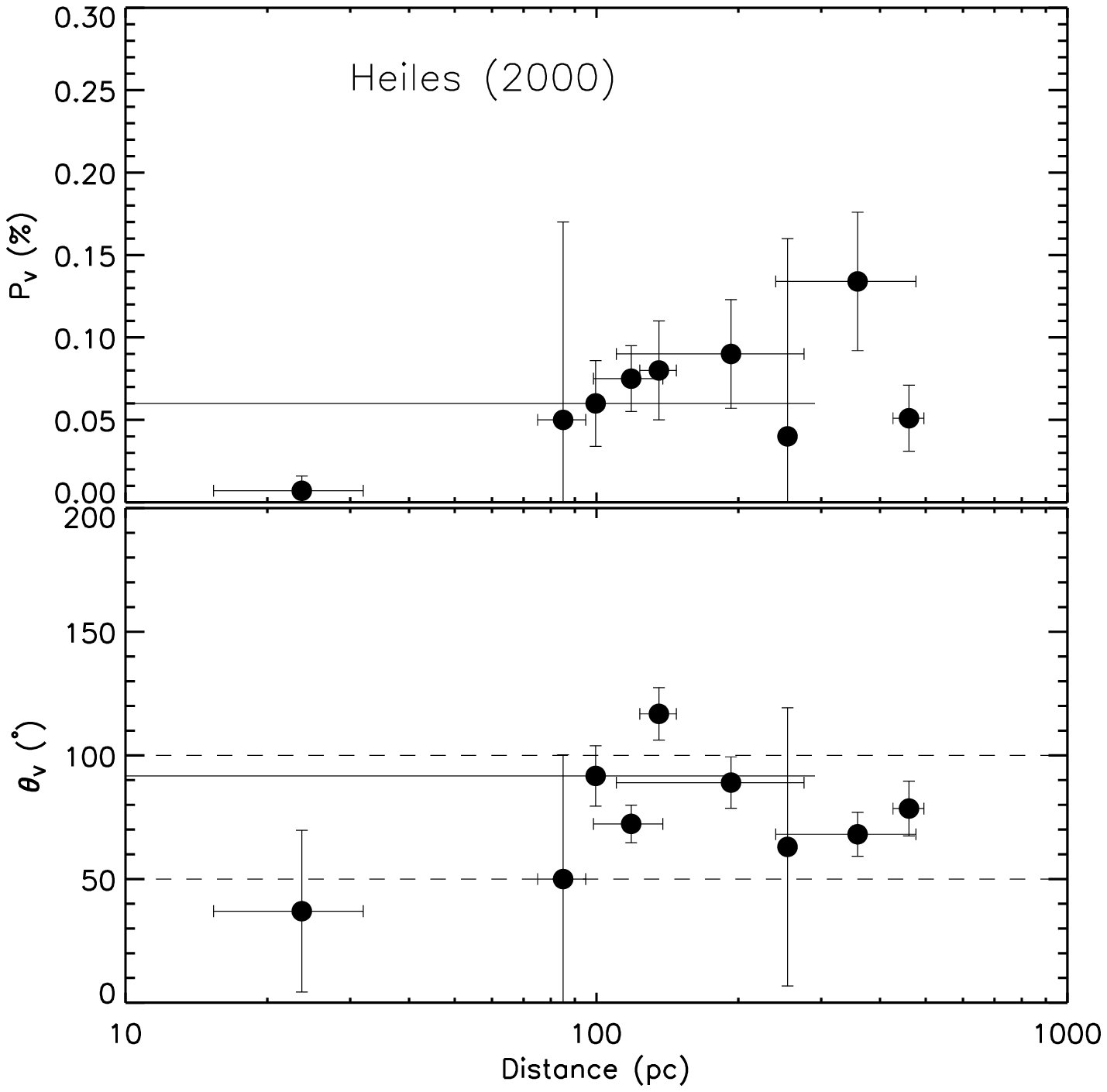}
\includegraphics[scale = 0.57]{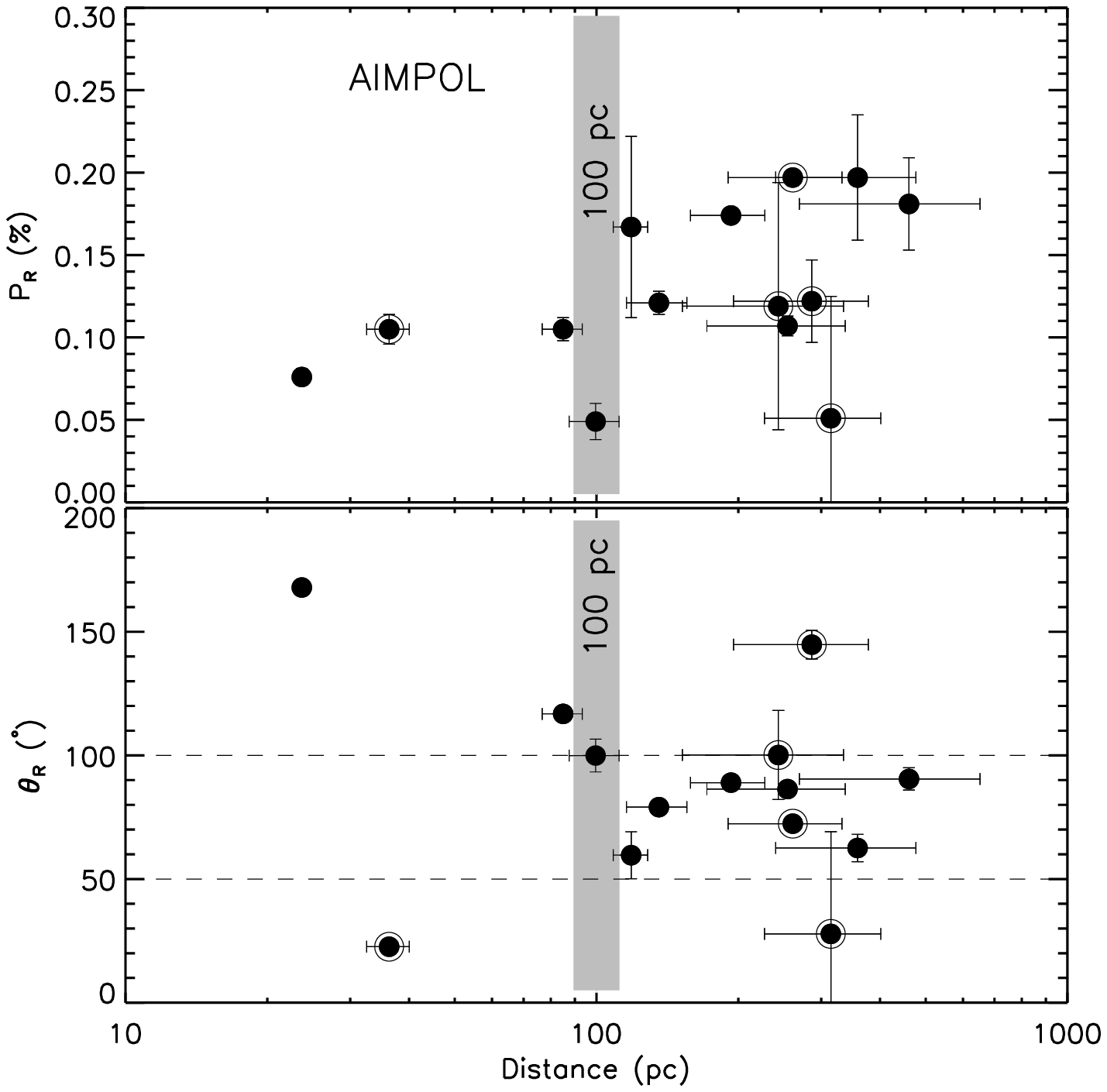}
\caption{Distribution of polarization and polarization angle of stars around SN~2012aw.
Left panel: 9 isolated field stars with known polarization and parallax measurements from 
\citet{2000AJ....119..923H} and \citet{2007A&A...474..653V}, respectively.
Right panel: same as left panel but for 14 isolated stars with $R$ band polarimetric data 
using AIMPOL and with distance from \citet{2007A&A...474..653V} catalogue. Filled circles 
denote 9 common stars in both left and right panels. The encircled filled circles are 5 
stars distributed within a 2$\degr$ radius around the location of SN~2012aw. 
The gray region represents the possible presence of a dust layer around 100 pc distance.}
\label{heiles_aimpol}
\end{centering}
\end{figure*}

\begin{table*}
\centering
\caption{Estimated polarimetric parameters for ISP$_{\rm MW}$ (see Section~\ref{isp_calc} for detail).
\label{sn_star_fit}}
\begin{tabular}{cccccc}
\hline \hline
 Number     & Distance & $<$$Q_{R} \pm \sigma_{Q_{R}}$$>$ & $<$$U_{R} \pm \sigma_{U_{R}}$$>$ & $<$$P_{R} \pm \sigma_{P_{R}}$$>$ & $<$$\theta{_R} \pm \sigma_{\theta{_R}}$$>$ \\
 of stars   & (pc)     &   ($\%$)                   &     ($\%$)                 & ($\%$)                     &  ($^\circ$)                          \\
\hline
  14$^{\#}$ & all distances & $-$ 0.101 $\pm$ 0.002      & 0.012 $\pm$ 0.002  & 0.102  $\pm$ 0.002 & 86.49 $\pm$ 0.54 \\ 
  10        & $>$ 100  & $-$ 0.154 $\pm$ 0.002      & 0.032 $\pm$ 0.002  & 0.157  $\pm$ 0.002 & 84.10 $\pm$ 0.43 \\
\hline
\end{tabular}\\  
$^{\#}$ All stars within 10$\degr$ radius around the SN. \\
\end{table*}


\subsection{Interstellar polarization due to Milky Way (ISP$_{\rm MW}$)}\label{isp_calc}

To estimate the interstellar polarization in the direction of SN~2012aw, we have performed $R$-band
polarimetric observations of 14 isolated and non-variable field stars (which do not show either emission 
features or variability flag in the SIMBAD database) distributed in a region of 10$^\circ$ radius 
around SN. All 14 stars have distance information from Hipparcos parallax \citep{2007A&A...474..653V}
and out of these, 9 stars have both polarization \citep{2000AJ....119..923H} and distance measurements.
In Fig.~\ref{heiles_aimpol} (left panels), we show the distribution of degree of polarization and 
polarization angles for these 9 stars. The weighted mean values of $P_V$ and $\theta_V$ of 8 out of 
these 9 stars (after excluding one star whose $P_V$ is 0.007\%) are found to be 0.071\% $\pm$ 0.010\% 
and 83$\degr$ $\pm$ 4$\degr$, respectively. 
Because our polarimetric observations are performed in the $R$-band therefore, to correct 
for ISP$_{\rm MW}$ component and to study the intrinsic behavior of SN, we have used 
polarization measurements of 14 field stars observed on 20 January 2013 in $R$-band. 
The distribution of $P_{R}$ and $\theta_R$ values of these stars is shown in right panels 
of Fig.~\ref{heiles_aimpol}. All the observed 14 stars are shown with filled circles. 
As revealed by both left and upper right panels of Fig.~\ref{heiles_aimpol}, the amount of degree 
of polarization shows an increasing trend with distance. 
It is worthwhile to note that, in the upper right panel, the degree of 
polarization ($P_{R}$) seems to show a sudden jump from $\sim$0.1\% at the distance of 
$\sim$100 pc to $\sim$0.2\% at a distance of $\sim$250 pc, thereby indicating the presence 
of a dust layer (shown with a gray region in Fig.~\ref{heiles_aimpol}) at $\sim$100 pc.
Whereas, the polarization angles of the stars from Heiles catalogue (left bottom panel) 
and those observed from the present set-up in $R$-band (except few stars) are distributed 
between $50\degr-100\degr$ as shown with the dashed lines (in Fig.~\ref{heiles_aimpol}). 
The Gaussian mean value of $\theta_{R}$ using 14 stars is found to be $\sim$82$^\circ$. 
This indicates the presence of a uniform dust layer towards the direction of SN2012AW, which 
nearly contributes $\sim$0.1\% to $\sim$0.2\% of polarization and having a mean magnetic 
field orientation $\sim$82$^\circ$. Therefore, we believe that most probably, 
the ISP$_{\rm MW}$ component has been dominated by the contribution from this dust layer.

To determine the ISP$_{\rm MW}$ component, firstly the $P_{R}$ and $\theta_{R}$ values
of all field stars as well as SN were transformed into Stokes parameters using the following 
relations\footnote{Our polarimeter and software have been designed in such a way that we get 
$P$ and $\theta$ through fitting the equation \ref{R_alpha} on four Stokes parameters obtained 
at four positions of half-wave plate as mentioned in Section~\ref{sec:obs_analy}}: 

\begin{equation}
Q_{R} = P_{R} \cos 2\theta_{R}
\end{equation}

\begin{equation}
U_{R} = P_{R} \sin 2\theta_{R}
\end{equation}

Then, the weighted mean Stokes parameters were estimated by considering (a) all 14 field 
stars distributed over all distances, and (b) only 10 field stars distributed beyond the 
distance of 100 pc. These weighted Stokes parameters ($<$$U_{R}$$>$, $<$$Q_{R}$$>$) were 
converted back to $P_{R}$ and $\theta_{R}$ using the following relations:

\begin{equation}\label{eq3}
P_{R} = \sqrt{{{Q_{R}}^2} + {{U_{R}}^2}}
\end{equation}

\begin{equation}\label{eq4}
\theta_{R} = 0.5 \times \arctan\left(\frac{U_{R}}{Q_{R}}\right)  \\ 
\end{equation}

The $<$$U_{R}$$>$, $<$$Q_{R}$$>$, $<$$P_{R}$$>$ and $<$$\theta_{R}$$>$ values 
(as estimated by two ways) are listed in Table~\ref{sn_star_fit}. It is clear from 
this table that the $<$$P_{R}$$>$ of 14 stars is relatively smaller than that determined 
using 10 stars. This could be due to the fact that the weighted mean values 
for stars at all distances may skew the result towards the brighter and more nearby
stars which likely to be incorrect. Whereas the $<$$\theta_{R}$$>$ values in 
two cases nearly matches with each other and mimic the mean magnetic field 
orientation ($\sim$82$^\circ$) of the dust layer as noticed above. 
To avoid the values biased towards lower end due to nearby and brighter stars, 
we have considered the polarization measurements of 10 stars distributed beyond 
100 pc distance to estimate the ISP$_{\rm MW}$ component. In addition, using these 10 stars 
which are distributed beyond 100 pc essentially may take care of the contribution 
from the dust layer at the distance of 100 pc. Therefore, we consider 
$<$$Q_{R}$$>$ = $-$ 0.154 $\pm$ 0.002\%, 
$<$$U_{R}$$>$ = 0.032 $\pm$ 0.002\% values as the ISP$_{\rm MW}$ component
(i.e. $<$$Q_{\it ISP_{\it MW}}$$>$ = $<$$Q_{R}$$>$ and $<$$U_{\it ISP_{\it MW}}$$>$ = $<$$U_{R}$$>$).
These weighted mean values have been subtracted vectorially from the Stokes 
parameters of the SN using the relations: 

\begin{equation}
Q_{int} = Q_{\it SN} - \textless Q_{\it ISP_{\it MW}} \textgreater \\ 
\end{equation}
\begin{equation}
U_{int} = U_{\it SN} - \textless U_{\it ISP_{\it MW}} \textgreater \\ 
\end{equation}

Where $Q_{\it SN}$, $U_{\it SN}$ and $Q_{int}$ , $U_{int}$ denote respectively the observed 
and intrinsic (ISP$_{\rm MW}$ corrected) Stokes parameters of the SN. The resultant 
intrinsic Stokes parameters ($Q_{int}$, $U_{int}$) were converted into $P_{int}$ and $\theta_{int}$ 
using the relations \ref{eq3} and \ref{eq4}. 
These intrinsic values of SN are respectively listed in column 6 and 7 in 
Table~\ref{sn2012aw_log} and plotted in Fig.~\ref{aimpol}(a)
and (b), with filled circles connected with a thick line.

The reddening, $E(B-V)$ due to Milky Way dust in the direction of SN 2012aw, as derived from 
the 100-$\mu$m all-sky dust extinction map of \citet*{1998ApJ...500..525S}, was found 
to be 0.0278 $\pm$ 0.0002 mag. According to the mean polarization efficiency relation 
$P_{mean}$ = 5 $\times$ $E(B-V)$ \citep{1975ApJ...196..261S}, the polarization value is 
estimated to be $P_{mean}$ $\sim$0.14\% which closely matches with the weighted 
mean polarization value, 0.157 $\pm$ 0.002\% obtained using the 10 fields stars distributed 
beyond 100 pc distance (cf. Table~\ref{sn_star_fit}). It is clear that polarization values 
obtained both from the present observations of the field stars and mean polarization 
efficiency relation are similar which implies that the dust grains in the local 
interstellar medium (ISM) exhibit mean polarization efficiency. 


\begin{figure}
\centering
\includegraphics[scale = 0.115]{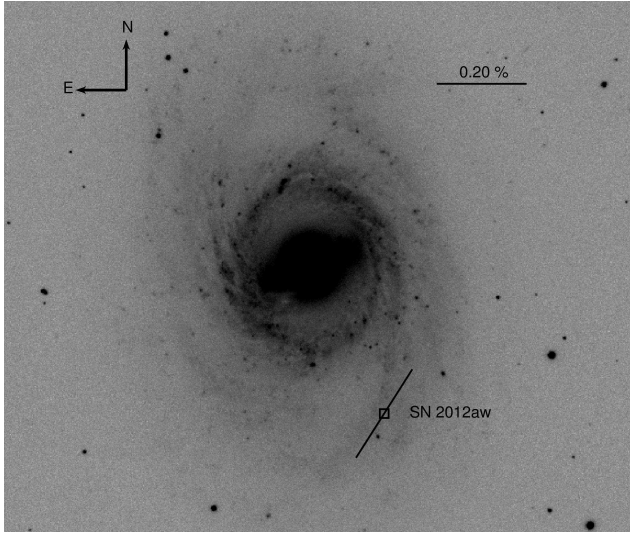}
\caption{{\sc SDSS} $g$-band image ($7\farcm7\times7\farcm2$) of the SN field containing 
the galaxy M95. A vector with a degree of polarization 0.23\% and position angle of 147$\degr$
is drawn at the location of SN 2012aw (see text in Section~\ref{igp_calc} for details).
A vector with a 0.20\% polarization and polarization angle of 90$\degr$ is shown for a 
reference (top right).
The approximate orientation of the magnetic field at the location of SN has been determined on the
basis of structure of the spiral arm (see Section~\ref{igp_calc} for more details).
The location of the SN is represented by a square symbol. North is up and east is to left as
shown in the figure.}
\label{sdsssn12aw_field}
\end{figure}

\subsection{Interstellar polarization due to host galactic dust (ISP$_{\rm HG}$)}\label{igp_calc}

The reddening, $E(B-V)$, due to dust in the SN~2012aw host galaxy was found to be 0.046 $\pm$ 0.008 
mag \citep[see][]{2013MNRAS.433.1871B}. This value was derived using the empirical correlation,
between reddening and Na{\sc~I} D lines, given by \citet{2012MNRAS.426.1465P}. 
As described in Section\ref{isp_calc}, the weighted mean value of polarization 
of 10 field stars situated beyond 100 pc distance (0.157\% $\pm$ 0.002\%) 
and the extinction (0.0278$\pm$0.0002 mag) due to the Galactic dust in the 
line of sight to the SN suggest that Galactic dust exhibits a mean polarization efficiency. 
To subtract the ISP$_{\rm HG}$ component we should estimate the degree of polarization and 
the magnetic field orientation of the host galaxy at the location of the SN. 

The properties of dust grains in the nearby galaxies have been investigated in detail only in 
a handful of cases and diverse nature of dust grains have been established in these studies. 
In case of SN~1986G, \citet{1987MNRAS.227P...1H} probed the ISP$_{\rm HG}$ component due 
to the dust lanes in the host galaxy NGC~5128 (Centaurus A) and validated that the size of the 
dust grains is smaller than typical Galactic dust grains. In another study (SN~2001el), the 
grains size were found to be smaller for NGC~1448 \citep{2003ApJ...591.1110W}. However, in some 
cases polarization efficiency of dust has been estimated to be much higher than the typical 
Galactic dust \citep[see e.g.][]{2002AJ....124.2506L, 2004AJ....127.3382C}.
For the present study, we assume that the dust grain properties of the M95 are similar 
to that the Galactic dust, and follow mean polarization efficiency relation 
\citetext{i.e. $P_{mean}$ = $5\times$$E(B-V)$; \citealt{1975ApJ...196..261S}}. Therefore, the
estimated polarization value would be $\sim$0.23\%. 

Another required parameter is the orientation of the magnetic field near the location of the SN. 
It is well known that large-scale Galactic magnetic field runs almost parallel (i.e. perpendicular 
to the line connecting a point with the galaxy center) to the spiral arms
\citep{1990IAUS..140..245S, 1991MNRAS.249P..16S, 1996ASPC...97..457H, 2009IAUS..259..455H}.
Interestingly, as shown in Fig.~\ref{sn12aw_field}, the SN~2012aw is located nearer to one of 
the spiral arms of the host galaxy. On the basis of the structure of the spiral arm and the 
location of SN, we have estimated the tangent to the spiral arm at the location of the SN (see 
Fig.~\ref{sdsssn12aw_field}), which makes approximately 147$\degr$ from the equatorial north 
increasing towards the east. We assume, on the basis of structure of the spiral arms and the 
magnetic field orientation that the magnetic field orientation in the host galaxy at the 
location of the SN is to be $\sim$147$\degr$. 
Here, we would like to emphasize that present procedure of considering magnetic field 
for the host of SN~2012aw is well established in previous spectropolarimetric studies
of Type IIP SN~1999em \citep{2001ApJ...553..861L} and Type IIb SN~2001ig 
\citep{2007ApJ...671.1944M}.   

\begin{figure}
\begin{centering}
\includegraphics[scale = 0.57]{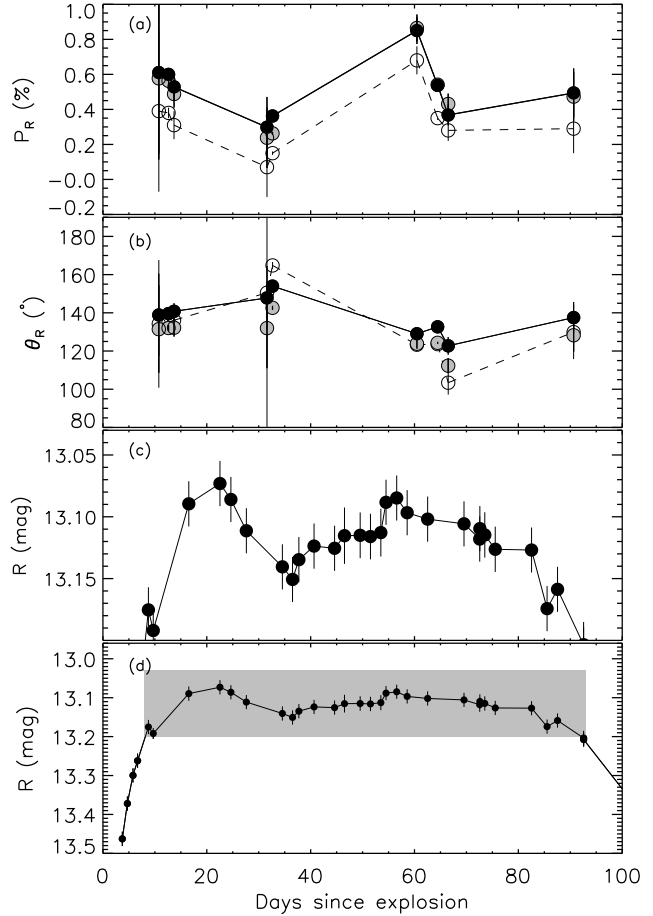}
\caption{Panels (a) and (b): Temporal evolution of polarization and polarization
angles of SN~2012aw in $R$ band, respectively. Filled circles connected with thick lines
denote the temporal evolution of polarization and polarization angles after subtracting
the ISP$_{\rm MW}$ component only, whereas those corrected for both
ISP$_{\rm MW}$ + ISP$_{\rm HG}$ components were represented with open circles
connected with broken lines. The observed polarization parameters are shown with gray
filled circles in panels (a) and (b). The bottom panel (d) shows the calibrated 
$R$ band LC of SN~2012aw obtained with ST \citep[see][]{2013MNRAS.433.1871B}. The photometric 
data shown with shaded region in the bottom panel (d) is re-plotted in panel (c) for a better
clarity.}
\label{aimpol}
\end{centering}
\end{figure}

As shown in the Fig.~\ref{sdsssn12aw_field}, a black vector with a length of 0.23\% and 
orientation of 147$\degr$ is drawn at the location of the SN which is shown with a square 
symbol. Hence, by assuming that the amount of polarization and the polarization angle due 
to the host galaxy are as 0.23\% and 147$\degr$, the Stokes parameters are estimated to be
$Q_{\it ISP_{\it HG}}$ = 0.11\%, $U_{\it ISP_{\it HG}}$ = $-$ 0.25\%. 
To get the intrinsic Stokes parameters and hence the amount of polarization and polarization 
angles purely due to the SN~2012aw, these values were subtracted vectorially from the 
ISP$_{\rm MW}$ corrected Stokes parameters as described in Section~\ref{isp_calc}. 
The intrinsic (ISP$_{\rm MW}$ + ISP$_{\rm HG}$ subtracted) polarization and polarization
angles of SN are listed in the Columns 8 and 9 of Table~\ref{sn2012aw_log} and plotted 
in the Fig.~\ref{aimpol} (a) and (b), respectively, with open circles connected with broken 
lines.

\begin{table*}
\centering
\caption{Observed and intrinsic (ISP$_{\rm MW}$ and ISP$_{\rm MW}$ + ISP$_{\rm HG}$ subtracted)
$Q-U$ parameters for SN~2012aw.
\label{sn2012aw_log_qu}}
\begin{tabular}{ccc|cc|cc|cc}
\hline \hline
UT Date &JD      & Phase$^{a}$& \multicolumn{2}{c}{Observed} & \multicolumn{2}{c}{Intrinsic (ISP$_{\rm MW}$} subtracted)& \multicolumn{2}{c}{Intrinsic (ISP$_{\rm MW}$ + ISP$_{\rm HG}$} subtracted) \\
(2012)  &2450000 &(Days)      & $Q_{R} \pm \sigma_{Q_{R}}$   & $ U{_R} \pm \sigma_{U{_R}}$ & $Q_{R} \pm \sigma_{Q_{R}}$ & $U{_R} \pm \sigma_{U{_R}}$ & $Q_{R} \pm \sigma_{Q_{R}}$ & $U_{R} \pm \sigma_{U_{R}}$ \\
        &        &            & ($\%$)            &($\%$) &   ($\%$) &  ($\%$)  &   ($\%$) &  ($\%$)                 \\
\hline
Mar 26 & 6013.35 & 10.75 & -0.072 $\pm$ 0.461 & -0.573 $\pm$ 0.464 &  0.082 $\pm$ 0.461 & -0.605 $\pm$ 0.464 & -0.012 $\pm$ 0.461 & -0.394 $\pm$ 0.464\\
Mar 28 & 6015.23 & 12.63 & -0.058 $\pm$ 0.029 & -0.560 $\pm$ 0.029 &  0.095 $\pm$ 0.029 & -0.592 $\pm$ 0.029 &  0.002 $\pm$ 0.029 & -0.382 $\pm$ 0.029\\
Mar 29 & 6016.28 & 13.68 & -0.048 $\pm$ 0.079 & -0.486 $\pm$ 0.078 &  0.106 $\pm$ 0.079 & -0.518 $\pm$ 0.078 &  0.012 $\pm$ 0.079 & -0.308 $\pm$ 0.078\\
Apr 16 & 6034.18 & 31.58 & -0.025 $\pm$ 0.174 & -0.237 $\pm$ 0.173 &  0.129 $\pm$ 0.174 & -0.269 $\pm$ 0.173 &  0.035 $\pm$ 0.174 & -0.059 $\pm$ 0.173\\
Apr 17 & 6035.25 & 32.65 &  0.069 $\pm$ 0.010 & -0.254 $\pm$ 0.009 &  0.223 $\pm$ 0.010 & -0.286 $\pm$ 0.009 &  0.129 $\pm$ 0.010 & -0.076 $\pm$ 0.009\\
May 15 & 6063.05 & 60.45 & -0.330 $\pm$ 0.077 & -0.800 $\pm$ 0.077 & -0.176 $\pm$ 0.077 & -0.832 $\pm$ 0.077 & -0.269 $\pm$ 0.077 & -0.622 $\pm$ 0.077\\
May 19 & 6067.04 & 64.44 & -0.198 $\pm$ 0.009 & -0.504 $\pm$ 0.009 & -0.044 $\pm$ 0.010 & -0.536 $\pm$ 0.009 & -0.137 $\pm$ 0.010 & -0.326 $\pm$ 0.009\\
May 21 & 6069.08 & 66.48 & -0.307 $\pm$ 0.059 & -0.303 $\pm$ 0.059 & -0.153 $\pm$ 0.059 & -0.335 $\pm$ 0.059 & -0.247 $\pm$ 0.059 & -0.125 $\pm$ 0.059\\
Jun 14 & 6093.23 & 90.63 & -0.111 $\pm$ 0.141 & -0.460 $\pm$ 0.140 &  0.043 $\pm$ 0.141 & -0.492 $\pm$ 0.140 & -0.050 $\pm$ 0.141 & -0.282 $\pm$ 0.140\\
\hline
\end{tabular}  \\
$^{a}$ with reference to the explosion epoch JD 2456002.6\\
\end{table*}

\section{Discussion}\label{diss}

\subsection{Polarization light curve (PLC) analysis}\label{pcl}

In this section, we analyze the evolution of the PLC and its possible resemblance 
with the photometric light curve (LC) of the SN~2012aw as shown in Fig~\ref{aimpol}. 
The calibrated $R$-band magnitudes have been taken from \citet{2013MNRAS.433.1871B} 
which shows different evolutionary phases of the LC as described in  
\citet{1971Ap&SS..10...28G, 1977ApJS...33..515F, 2007A&A...461..233U}.  
Since in the present study, polarimetric data sets are limited upto the plateau 
phase, in Fig.~\ref{aimpol} (panels c and d), only the adiabatic cooling phase 
and the phase of cooling and recombination wave are shown.  

The temporal variation of ISP$_{\rm MW}$ corrected degree of polarization ($P_{R}$) 
values (shown with filled circles, Fig.~\ref{aimpol}a) shows a maximum and minimum values of 
$\sim$0.9\% and $\sim$0.3\%, respectively with a possible trend of variations in 
accordance with the $R$-band LC as shown in the panel \ref{aimpol}(c). Although there is a significant 
reduction in ISP$_{\rm MW}$ + ISP$_{\rm HG}$ corrected $P_{R}$ values (open circles, 
Fig.~\ref{aimpol}a), its resemblance with photometric light curve (panel c) remain similar. 
However, both ISP$_{\rm MW}$ and ISP$_{\rm MW}$ + ISP$_{\rm HG}$ corrected polarization 
angles ($\theta_{R}$, shown with filled circles in Fig.~\ref{aimpol}(b)) does not show much 
variation during the similar epochs of observations and are distributed around a weighted mean value 
of $\sim$138$\degr$. Interestingly, first (10-14 days) three measurements of ISP$_{\rm MW}$ 
corrected $P_{R}$ and $\theta_{R}$ are almost constant. During this adiabatic cooling phase, 
the SN LC seems to be brightened by $\sim$0.12 magnitude as shown in the 
Fig.~\ref{aimpol}c. 

It is worthwhile to note that dips observed around 35 days in the LC of the SN and in the 
ISP$_{\rm MW}$ + ISP$_{\rm HG}$ corrected $P_{R}$ are temporally correlated with a 
minimum amount of polarization ($\sim$0.07\%). This observed feature during the end of the 
adiabatic cooling or early recombination phase could be attributed to several reasons e.g., 
(i) changes in the geometry i.e., transition from more asphericity to sphericity of SN,
(ii) modification in density of scatterers (electrons and/or ions),
(iii) mechanism of scattering i.e., single and (or) multiple scattering,
(iv) changes in the clumpiness in the SN envelope,
(v) changes in the electron-scattering atmosphere of SN,
and (vi) interaction of SN with a dense circumstellar medium.
In the recombination phase ($\sim$40 days onwards), the evolution in the values of 
ISP$_{\rm MW}$ + ISP$_{\rm HG}$ corrected $P_{R}$ and $\theta_{R}$ are in such a way 
that the amount of polarization shows an increasing trend. This increasing trend could suggest 
that during the recombination phase and onwards, the geometry of the SN envelope could have 
acquired more asphericity. 

If we assume that the ISP$_{\rm MW}$ and ISP$_{\rm HG}$ components are constant, then 
the changes observed in the temporal variation of intrinsic polarization measurements of the SN 
could purely be attributed to variations in the geometry of the SN along with the other possible 
reasons such as the interaction of the SN shock with the ambient medium. However, these properties 
could be well addressed using high resolution spectroscopic/spectropolarimetric investigations 
which are beyond the scope of this paper.

\begin{figure*}
\begin{centering}
\includegraphics[scale = 0.46]{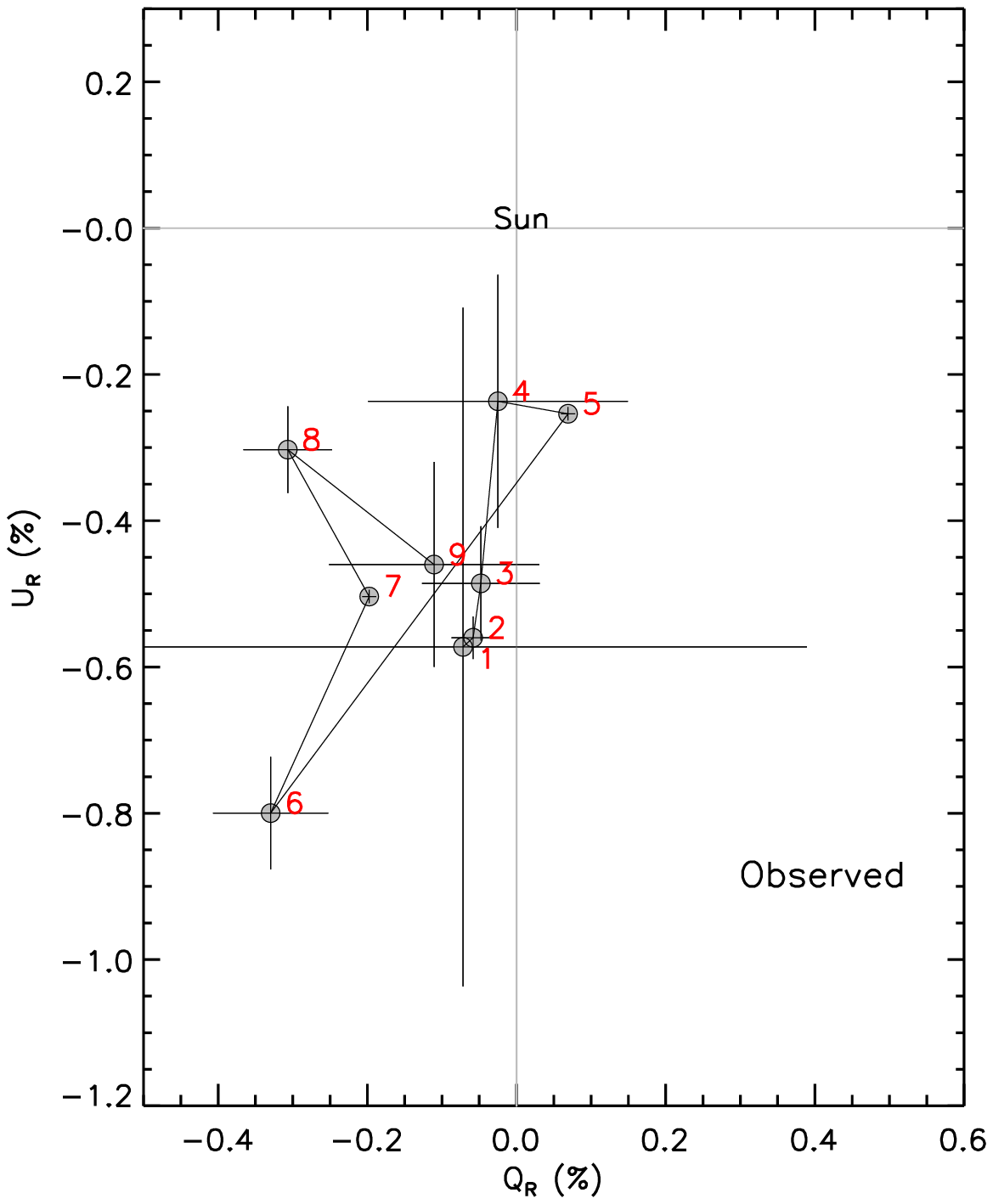}
\includegraphics[scale = 0.46]{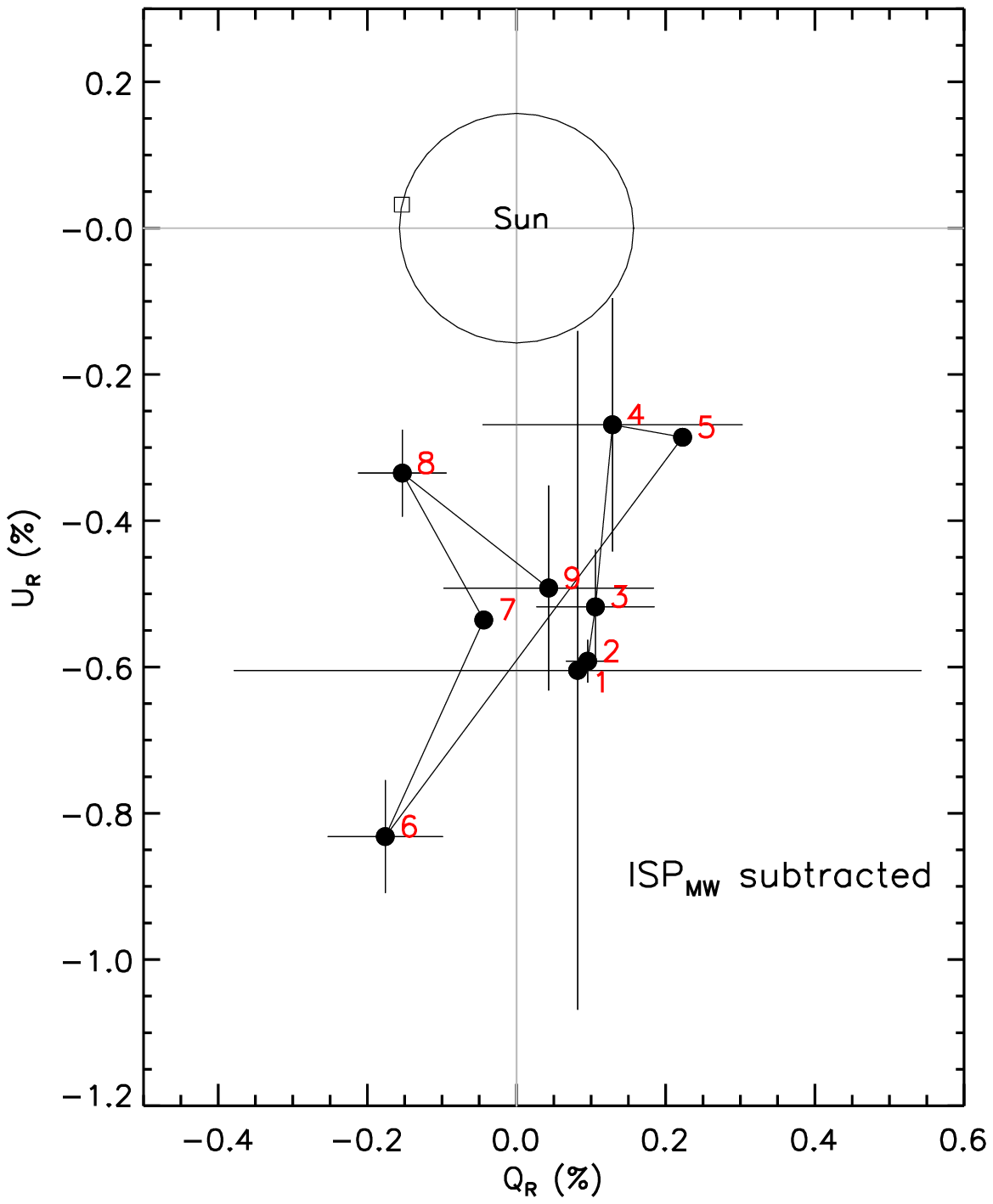}
\includegraphics[scale = 0.46]{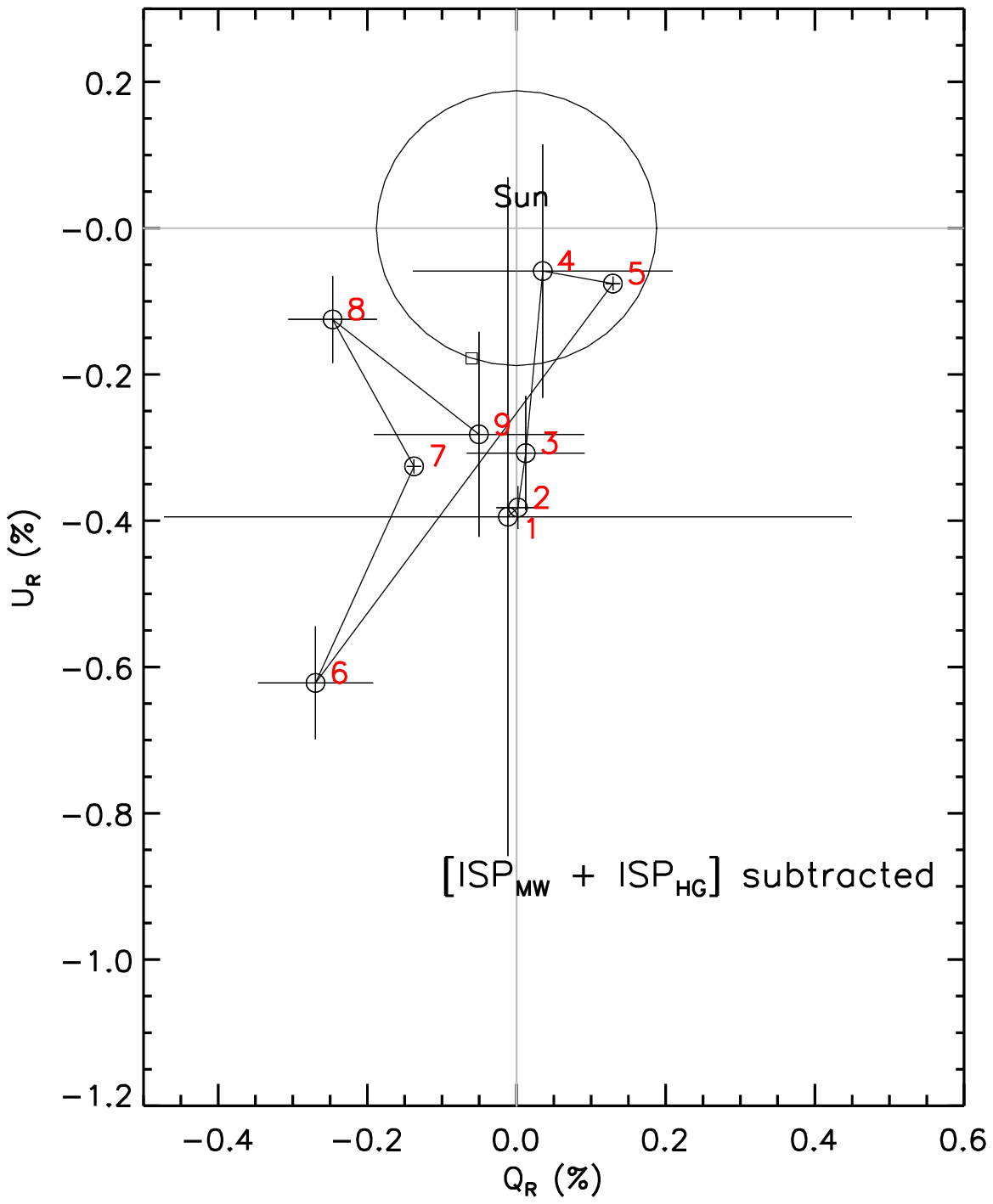}
\caption{Stokes $Q$ and $U$ parameters of SN 2012aw. Left panel: Gray filled circles are 
the observed parameters. Middle panel: The data have been corrected for the ISP$_{\rm MW}$
component only (black filled circle; see text). Right panel: After correcting
both ISP$_{\rm MW}$ + ISP$_{\rm HG}$ components (open circle; see text).
The square symbol connected with large circles drawn nearer to the solar neighborhood
in the middle and right panels, respectively, indicate the ISP$_{\rm MW}$
and ISP$_{\rm MW}$ + ISP$_{\rm HG}$ components.
Numbers labelled with 1 to 9 (red colour) and connected with continuous lines, indicate 
the temporal order.}
\label{qu}
\end{centering}
\end{figure*}

\subsection{$Q$ and $U$ parameters}\label{qup}

The $Q-U$ parameters, representing different projections of the polarization vectors, are 
used as a powerful tool to examine the simultaneous behavior of the polarization and the 
polarization angle with wavelengths \citep[see e.g.][]{2003ApJ...591.1110W, 2003ApJ...592..457W}. 
The pattern of the variation in $Q-U$ plane does not depend upon the 
ISP$_{\rm MW}$/ISP$_{\rm HG}$ corrections. 
However, the ISP$_{\rm MW}$/ISP$_{\rm HG}$ 
subtracted parameters are dependent on the corrections applied to the observed values. A small 
change in ISP$_{\rm MW}$/ISP$_{\rm HG}$ may considerably affect the polarization angle 
($PA$) values. 

The estimated $Q-U$ parameters (observed and intrinsic) for SN~2012aw are presented
in Table~\ref{sn2012aw_log_qu} and are plotted in Fig.~\ref{qu}.
The left and middle panels of this figure show the observed and ISP$_{\rm MW}$ 
subtracted parameters and, the right panel represents intrinsic parameters after 
subtracting both ISP$_{\rm MW}$ + ISP$_{\rm HG}$ contribution as discussed in 
Section~\ref{sec:results_discussion}.
The square symbol connected with large circles drawn nearer to the solar neighborhood 
in the middle and right panels respectively indicate 
ISP$_{\rm MW}$ ($Q_{\it ISP_{\it MW}}$ = $-$ 0.154, 
$U_{\it ISP_{\it MW}}$ = 0.032) and ISP$_{\rm MW}$ + ISP$_{\rm HG}$
($Q_{\it ISP_{\it MW} + ISP_{\it HG}}$ = $-$ 0.060, 
$U_{\it ISP_{\it MW} + ISP_{\it HG}}$ = $-$ 0.178)
components. 

Since, in the present case, the data points are limited, a firm conclusion could not be 
robustly drawn on behalf of $Q$ and $U$ parameters.
However, it seems that in all three panels of Fig.~\ref{qu}, these data points show scattered 
distribution, which seems to form a loop like structure on $Q-U$ plane. This kind of structure has 
been also observed in case of SN~1987A \citep{1988MNRAS.231..695C}, SN~2004dj 
\citep{2006Natur.440..505L} and SN~2005af \citep{2006A&A...454..827P}. Although, it is to 
be noted that if we ignore one of the data points (observed on 21 May), the variation of 
$Q-U$ parameters will more likely to follow straight line and in this case the 
previous interpretation may not be true.

\begin{figure*}
\begin{centering}
\includegraphics[width=15cm, height=15cm]{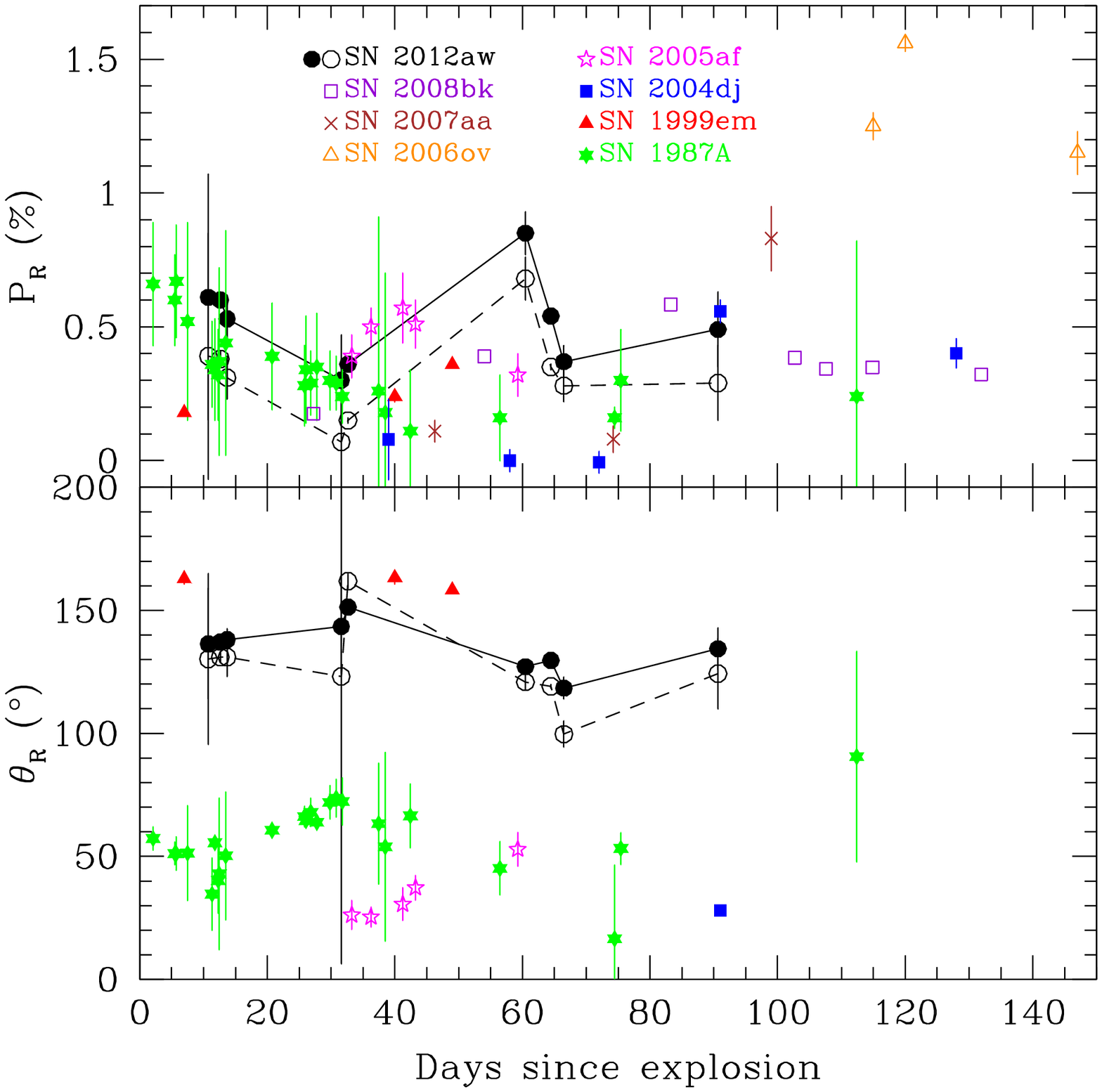}
\caption{Comparison of polarization and polarization angle values of SN~2012aw with those of 
other Type IIP SNe: SN~1987A, SN~1999em, SN~2004dj, SN~2005af, SN~2006ov, SN~2007aa and SN~2008bk.
The upper and lower panels show the degree of polarization and polarization angle, respectively.
All values are intrinsic to a particular SN and symbols used in both panels are same. 
Thick and broken lines denote ISP$_{\rm MW}$ and both ISP$_{\rm MW}$ + ISP$_{\rm HG}$ 
subtracted components, respectively for SN~2012aw.}
\label{comp}
\end{centering}
\end{figure*}

\subsection{Comparison with other Type IIP events}

We have collected polarization parameters of a few well-observed Type IIP SNe from literature: 
SN~2008bk \citep{2012AIPC.1429..204L}, SN~2007aa and SN~2006ov \citep{2010ApJ...713.1363C}, 2005af 
\citep{2006A&A...454..827P} 2004dj \citep{2006Natur.440..505L}, 1999em \citep{2001ApJ...553..861L} 
and SN~1987A \citep{1988MNRAS.234..937B} for which polarimetric observations have been performed for 
two or more epochs. Except for SN~1987A, SN~2005af and SN~2012aw, the data for other events are 
spectropolarimetric only. The intrinsic polarization values of SN~2012aw along with other SNe are 
plotted in Fig.~\ref{comp}. It is worthwhile to note that the explosion epochs of SN~1987A 
\citep[see][]{1987PhRvL..58.1490H, 1987PhRvL..58.1494B}, SN~1999em~\citep[see][]{2003MNRAS.338..939E}
and SN~2012aw are known precisely, but there is uncertainty in the estimation of explosion epoch 
for other events (SN~2004dj, SN~2005af, SN~2006ov, SN~2007aa and SN~2008bk). In case of SN~2004dj, 
\citet{2006Natur.440..505L} considered the explosion epoch on JD 2453200.5 but 
\citet{2006AJ....131.2245Z} estimated it on JD 2453167 $\pm$ 21. With an uncertainty of few weeks, 
the explosion epoch for SN~2005af is estimated to be on JD 2453379.5 \citep[see][]{2006ApJ...651L.117K}.
For SN~2006ov, \citet{2006CBET..757....1B} estimated the expected date of explosion $\sim$36 days 
before the discovery \citep{2006CBET..756....1N} but \citet{2007ApJ...661.1013L} reasonably constraints 
its explosion about 3 months before the discovery. We follow the later study in the present analysis. 
Similarly we considered explosion epoch for SN~2007aa, $\sim$20 days before the discovery 
\citep[see][]{2007CBET..848....1D, 2007CBET..850....1F} and for SN~2008bk, JD 2454550 (2008 March 24)
has been considered as explosion epoch \citep[see][]{2008CBET.1335....1M, 2010arXiv1011.5873V}. 


Shifting the phase (days after the explosion) by 21 days, the evolution of degree of polarization 
of SN~2004dj is very much similar to what has been seen in case of SN~2012aw as shown in the 
Fig.~\ref{comp}. However, it is important to mention that in case of SN~2004dj
the degree of polarization increases after the end of the plateau phase (when we see through the 
H-rich shell); whereas for SN~2012aw, the degree of polarization increases (around 60 days) during 
the plateau phase which could be a possible indication of a diverse nature of the two events. 
However, it is noticeable that like SN~2008bk, SN~2012aw is also strongly polarized well before the 
end of the plateau \citep[see][]{2012AIPC.1429..204L}, indicating a possible similarity in the both 
events. In the early phase ($\sim$10 $-$ 30 days), the ISP$_{\rm MW}$ corrected PLC of 
SN~2012aw matches with that of the SN~1987A, whereas in the later phase ($\sim$30 $-$ 45 days) 
it is closely matching with that of SN~2005af. Nonetheless, it is worthwhile to mention that 
to derive the polarization parameters of SN~1987A and SN~2005af, the ISP$_{\rm HG}$ 
components are not subtracted in respective studies. The polarimetric observations of SN~1999em 
are sparse; the polarization levels at different epochs seems to match with the 
ISP$_{\rm MW}$ + ISP$_{\rm HG}$ corrected PLC of SN~2012aw.  
It is also obvious from figure \ref{comp} that the polarization values of SN~2006ov remains more than
1\% for all three epoch observations which is higher than that of any of the Type IIP events in the 
list. Fig.~\ref{comp} gives an important information regarding the evolution of ejecta of similar 
types of SNe. By comparing the PLCs of various IIP SNe shown with different symbols in Fig.~\ref{comp} 
(filled star: SN~1987A, filled triangle: SN~1999em, filled square: SN~2004dj, open star: SN~2005af, open 
triangle: SN~2006ov, cross: SN~2007aa, open square: SN~2008bk and for SN~2012aw symbols are the
same as in Fig.~\ref{aimpol}), it could be conjectured that the properties of the ejecta from 
Type IIP SNe are diverse in nature as noticed by \citet{2010ApJ...713.1363C}. 

We have also compared the ISP$_{\rm MW}$ and ISP$_{\rm HG}$ corrected PLCs of 
SN~2012aw with those of other Type Ib/c CCSNe. Type Ib/c SNe are naturally more asymmetric 
in comparison to that of Type IIP SNe because they lack a thick He blanket that smear out 
the internal geometry. Therefore, a higher degree of polarization is observed in case 
of Type Ib/c SNe. In the present analysis, PLC of SN~2012aw is also clearly showing a 
lower degree of polarization in comparison to well studied various Type Ib/c CCSNe 
\citep[e.g. SN~2007uy, SN~2008D;][]{2010A&A...522A..14G}. However, it is important 
to note that the $P_{R}$ peak value for SN~2012aw seen at $\sim$60 days is slightly less 
than the intrinsic polarization value of $\sim$1\% for the Type Ic SN~2008D which was 
related to violent X-ray transient \citep[see][]{2010A&A...522A..14G}.
Here it is noticeable that present PLC interpretations of SN~2012aw depend
a lot on a single data point (May 15) which is significantly higher in the percentage 
polarization than the data taken at other epochs.

\section{Conclusions}\label{sec:conclusions}

We present results based on 9 epoch $R$ band imaging polarimetric observations of Type
IIP supernova SN~2012aw. To the best of our knowledge, the initial three epoch polarimetric 
observations presented here are the earliest optical polarimetric data reported for this event. 
It was not possible to monitor the SN during the beginning of the nebular or post-nebular 
phase due to observational constraints, however present observations cover almost up to 
the end of the plateau phase ($\sim$90 days).
The main results of our present study are the following:

\begin{itemize}  
  
\item {The observed broad-band polarization for initial three epochs is $\sim$0.6\%, then
decreases up to $\sim$0.3\% following a sudden increase up to $\sim$0.9\% on 15 may 2013
and at later epochs it seems to show a declining trend. However, the observed polarization 
angle is almost constant, superimposed with slight variations.}

\item {To study the intrinsic polarization properties of SN~2012aw, we subtracted the
contribution due to ISP$_{\rm MW}$ and ISP$_{\rm HG}$ from the observed 
$P$ and $\theta$ values of SN. 
The ISP$_{\rm MW}$ component was determined using the polarimetric observations of 10 field 
stars distributed within 10$\degr$ radius around SN and located beyond 100 pc distance. 
The estimated Stokes parameters of ISP$_{\rm MW}$ are found to be
$<$$Q_{\it ISP_{\it MW}}$$>$ = $-$ 0.154 $\pm$ 0.002\% and
$<$$U_{\it ISP_{\it MW}}$$>$ = 0.032 $\pm$ 0.002\%
(equivalent to $<$$P_{\it ISP_{\it MW}}$$>$ = 0.157 $\pm$ 0.002 and
$<$$\theta_{\it ISP_{\it MW}}$$>$ = 84.10$\degr$ $\pm$ 0.56$\degr$). 
We also estimated the degree of polarization (0.23\%) and polarization angle (147$\degr$) 
at the location of SN by using the extinction value from Schlegel map assuming that 
the host galactic dust follow the mean polarization efficiency and the magnetic field in 
the host galaxy follow the structure of the spiral arms.}

\item {The intrinsic polarization parameters of SN~2012aw follow trends of the 
photometric LC which could be attributed to the small scale variations in the SN atmosphere 
or their interaction with the ambient medium.} 

\item{Polarimetric parameters of this SN are compared with other well studied Type IIP events. 
During the early phase ($\sim$10 $-$ 30 days), the ISP$_{\rm MW}$ subtracted PLC of 
SN~2012aw matches with that of SN~1987A whereas at later epochs ($\sim$30 $-$ 45 days) it 
matches to that of SN~2005af.}

\end{itemize}

\section*{Acknowledgments}
We are grateful to the observers Archana Soam, Manoj Kumar Patel and Ram Kesh Yadav
at the Aryabhatta Research Institute of observational sciencES (ARIES) for their valuable 
time and support for the observations of this event. SBP and BK acknowledge the support 
of the Indo-Russian (DST-RFBR) project No. INT/RFBR/P-100 for this work. SBP also
acknowledge Dr. Koji S. Kawabata for useful discussions on various polarimetric 
aspects. C. Eswaraiah  acknowledges the financial support of the Pan-STARRS grant 
NSC 102-2119-M-008-001 funded by the Ministry of Science and Technology of Taiwan
J. Gorosabel acknowledges support of the Unidad Asociada IAA/CSIC-UPV/EHU 
and the Ikerbasque science foundation. This work has been supported by Spanish 
Junta de Andaluc\'{\i}a through program FQM-02192 and from the Spanish 
Ministry of Science and Innovation through Projects (including FEDER funds) 
AYA 2009-14000-C03-01 and AYA2008-03467/ESP. 
This research has made use of the SIMBAD database, operated at CDS, Strasbourg, 
France. We acknowledge the usage of the HyperLeda database (http://leda.univ-lyon1.fr).

\label{lastpage}
\bibliography{draftsn12aw}
\end{document}